\documentclass[onecolumn,useAMS,usenatbib]{mn2e}
\usepackage{graphicx}% Include figure files
\usepackage{subfigure}% Include subfigure files
\usepackage{xcolor}% Change colors of text using \textcolor{red}{...}
\usepackage{mathtext,bbm,amsmath,amsfonts,amssymb,indentfirst,syntonly,graphicx}
\usepackage{slashbox}
\usepackage[english]{babel}
\usepackage{calc}
\usepackage{tikz}
\usepackage{bm}
\usepackage{epstopdf}

\def\bc{\begin{center}}
\def\ec{\end{center}}
\def\be{\begin{eqnarray}}
\def\ee{\end{eqnarray}}

\title[Testing the anisotropy of the Universe with DDR]{Testing the anisotropy of Universe with the distance duality relation}
\author[X. Li, L. Tang and H.-N. Lin]
        {Xin Li$^{1}$\thanks{E-mail: lixin1981@cqu.edu.cn},
        Li Tang$^{1}$\thanks{E-mail: tang@cqu.edu.cn} and
        Hai-Nan Lin$^{1}$\thanks{E-mail: linhn@ihep.ac.cn}\\
$^{1}$Department of Physics, Chongqing University, Chongqing 401331, China\\}
\begin{document}

\date{Accepted xxxx; Received xxxx; in original form xxxx}

\pagerange{\pageref{firstpage}--\pageref{lastpage}} \pubyear{2017}

\maketitle

\label{firstpage}

\begin{abstract}
The distance duality relation (DDR) is valid in Riemannian spacetime. The astronomical data hint that the universe may have certain preferred direction. If the universe is described by anisotropic cosmological models based on Riemannian spacetime, then DDR still valid. If the anisotropy universe is described by other models which are not based on Riemannian spacetime, then DDR is violated. Thus, DDR could be used to test the validity of these anisotropic cosmological models. In this paper, we perform anisotropic DDR parametrization with the dipolar structures. The DDR is tested by comparing the luminosity distance from type-Ia supernovae (Union 2.1 and JLA compilations) and the angular diameter distance from strong gravitational lensing (SL) systems at the same redshift. It is shown that, the DDR is valid with the Union2.1 compilation, while is violated more than 1$\sigma$ confidence level with the JLA compilation. Additionally, we verify the statistical signification of our method with Monte Carlo simulations. Due to the large uncertainty of available data, no strong evidence is found to violate the DDR in the anisotropic models.
\end{abstract}

\begin{keywords}
cosmology: distance scale -- galaxies: clusters: general -- supernovae: general
\end{keywords}

\section{Introduction}
\label{intro}
The relation between luminosity distance $D_{L}(z)$ and angular diameter distance $D_{A}(z)$ at a given redshift $z$, i.e. the so-called distance duality relation (DDR), has aroused enormous interests in astronomy and cosmology, since it was first introduced by \citet{Etherington:1933,Etherington:2007}. In the standard cosmological model, the DDR takes the simple form $D_{A}(z)(1+z)^{2}/D_{L}(z)=1$. \citet{Ellis:1971} proved this equation and showed that the DDR relation holds true in the conditions that, a) the spacetime is depicted by a metric theory of gravity, b) the photons travel along null geodesics and the number of photons is conserved. There will be new physics beyond the standard model if the DDR is violated. The violation of DDR could be induced by coupling of photon with particles beyond the standard model of particle physics \citep{Bassett:2004}, dust extinction \citep{Corasaniti:2006}, or varying fundamental constants \citep{Ellis:2013}. Hence, numerous works have been devoted to testing the validity of DDR. The most popular way is to compare the luminosity distance estimated from type-Ia supernovae (SNe Ia) and angular diameter distance estimated from strong gravitational lensing (SL) or galaxy clusters at the same redshift \citep{Uzan:2004,Holanda:2010,Holanda:2012,Piorkowska:2011,Yang:2013,Holanda:2016,Liao:2016}. Up to now, no strong evidence for the violation of DDR has been found. Therefore, the DDR has been used to probed the gas mass density profiles \citep{Cao:2016} and the shapes of galaxy cluster \citep{Holanda:2011}.

The standard model of cosmology, namely the $\Lambda$CDM model is based on the cosmological principle that our universe is homogeneous and isotropic in large scale. Although the $\Lambda$CDM model is well compatible with accurate astronomical data from the Wilkinson Microwave Anisotropy Probe \citep{Bennett et al.:2013,Hinshaw et al.:2013} and the Planck satellite \citep{Planck Collaboration XVI:2014a,Planck Collaboration XIII:2016}, it also encounters many challenges. Recently, it was found that the electromagnetic fine-structure constant varies with cosmological distance from earth by the analysis of a large sample of quasar absorption-line spectra \citep{Webb:2011,King:2012}. Additionally, the SNe Ia data hint that the universe may have a certain preferred direction \citep{Watkins:2009,Antoniou:2010,Chang:2013}. The recent Planck data show that the CMB temperature map possesses power asymmetry \citep{Ade:2014,Ade:2016}. All above indicate the possible violation of the cosmological isotropy, and motivate physicists to propose new models to describe the cosmology, such as the Bianchi cosmology \citep{Collins:1973,Barrow:1985,Pontzen:2009,Campanelli:2011}, the Finsler cosmology \citep{Chang:2013b,Chang:2014,Chang:2015,Li:2015}, the $\Lambda$CDM with a scalar perturbation \citep{Li:2013}.

\citet{Etherington:2007} have proved that the DDR is valid in Riemannian spacetime. Therefore, cosmological models based on Riemannian geometry, such as the Bianchi cosmological model, preserve the DDR. However, if the components of cosmology couples to photon, such as interaction between dark energy and photon \citep{Ellis:2013}, or the cosmology is depicted by a non Riemannian spacetime, such as the Finsler spacetime \citep{Li:2015}, then the DDR is violated in these cosmological models.
Thus, the DDR could be used to test the validity of these anisotropic cosmological models.

At present, the tests of DDR usually involve the measurement of $D_L$ and $D_A$ from two different objects locating at the same redshift. For example, $D_L$ can be extracted from SNe Ia, and $D_A$ can be obtained from strong gravitational lensing (SL). However, the two objects usually locate in different direction in the sky, so could not be compared directly if the universe is anisotropic. Inspired by these, in this paper we try to investigate the effect of anisotropy on the DDR. We parameterize the DDR similarly to previous works \citep{Holanda:2010,Liao:2016} but with the form involving the direction: $D_{A}(z)(1+z)^{2}/D_{L}(z)=1+A \cos\theta$, where $\theta$ represents the angle between two objects, $A$ is the amplitude of anisotropy, respectively. The SL sample used in this work is compiled in \citet{Cao:2015}. We also add galaxy clusters to SL systems to enlarge the sample. For SNe Ia sample, we use two different datasets, i.e. Union2.1 \citep{Suzuki:2012} and JLA \citep{Betoule:2014}. \citet{Lin:2016} searched for the anisotropic signal in two compilations of SNe Ia and found that the results are not consistent. Thus, it is interesting to test the DDR using two different SNe Ia compilations.

The rest of the paper is organized as follows: In Section \ref{sec:Observation and Methodology}, we introduce a dipolar anisotropic parametrization of DDR and describe the methodology to constrain the anisotropic parameters. In Section \ref{sec:Samples and Results}, we combine the SNe Ia, SL systems and galaxy clusters to give a constraint on the anisotropic parameters. Then we test our method with the Monte Carlo (MC) simulation in Section \ref{sec:Simulation}. Finally, discussion and summary are given in Section \ref{sec:summary}.

\section{Theory and Methodology}\label{sec:Observation and Methodology}

The main idea of testing the DDR is to compare the luminosity distance $D_L$ and angular diameter distance $D_A$ at the same redshift. If we can measure  both $D_L$ and $D_A$ of a specific object, we can compare them directly. However, it is difficult to measure both $D_L$ and $D_A$. In practice, $D_L$ and $D_A$ are usually measured from different kinds of objects. For example, it is easy to measure $D_L$ from SNe Ia, and $D_A$ from gravitational lensing.

Strong gravitational lensing plays a significant role in constraining cosmological parameters \citep{Treu:2006,Biesiada:2010}, testing cosmology models \citep{Zhu:2000,Mitchell:2005,Linder:2016} and the structure of galaxy cluster \citep{Yang:2013,Holanda:2011}. The Einstein radius ($\theta_{E}$) is a characteristic angle for gravitational lensing, which depends on the angular diameter distances between the lens and source $D_{A_{ls}}$, and between the observer and source $D_{A_{s}}$. In a singular isothermal sphere (SIS) lens model, Einstein radius can be written as
\begin{equation}\label{eq:E}
\theta_{E}=4 \pi \frac{D_{A_{ls}}}{D_{A_{s}}} \frac{\sigma^{2}_{\rm SIS}}{c^{2}},
\end{equation}
where $\sigma_{\rm SIS}$ is the velocity dispersion due to lens mass profile, and $c$ is the speed of light. \citet{White:1996} pointed out that $\sigma_{\rm SIS}$ does not necessary equal to the observed stellar velocity dispersion $\sigma_0$. Therefore, a phenomenological parameter $f$ is introduced to account for the difference between these two velocity dispersions, i.e. $\sigma_{\rm SIS} = f \sigma_{0}$ \citep{Kochanek:1992,Ofek:2003,Cao:2012}.

From equation (\ref{eq:E}), we could not obtain $D_{A_{ls}}$ and $D_{A_{s}}$ separately. However, we can derive the ratio of $D_{A_{ls}}$ and $D_{A_{s}}$ if both $\theta_E$ and $\sigma_{\rm SIS}$ are measured, i.e.,
\begin{equation}\label{eq:RA}
R_{A}\equiv\frac{D_{A_{ls}}}{D_{A_{s}}}=\frac{c^{2} \theta_{E}}{4\pi \sigma^{2}_{\rm SIS}}.
\end{equation}
The uncertainty of $R_{A}$ is propagated from that of $\theta_E$ and $\sigma_{\rm SIS}$,
\begin{equation}\label{eq:delta_RA}
\triangle R_{A}= R_{A}\sqrt{4\left(\frac{\triangle \sigma_{\rm SIS}}{\sigma_{\rm SIS}}\right)^{2}+\left(\frac{\triangle \theta_{E}}{\theta_{E}}\right)^{2}}.
\end{equation}

Due to the approximately constant absolute luminosity, SNe Ia are usually used as the distance indicators in cosmology. The luminosity distance can be extracted directly from the light curve of SNe Ia. The distance modulus of a SN Ia at redshift $z$ is given by
\begin{equation}\label{eq:mu}
\mu_{B}(z;\alpha,\beta,M_{B})=5{\rm log}_{10}D_{L}(z)+25=m_{B}-M_{B}+\alpha x(z)-\beta c(z),
\end{equation}
where $D_{L}(z)$ is the luminosity distance in unit of Mpc, $m_{B}$ is the apparent magnitude observed in rest frame B band, $x$ and $c$ are the stretch factor and color parameter, respectively. $M_{B}$, $\alpha$ and $\beta$ are nuisance parameters which can be derived using the least-$\chi^2$ method or be marginalized.
In a flat universe, the relation between the comoving distance $r(z)$ and angular diameter distance $D_{A}(z)$ is given by
\begin{equation}\label{eq:r}
r(z)=(1+z)D_{A}(z),
\end{equation}
and the comoving distance from lens to source is simplified to $r_{ls}=r_{s}-r_{l}$ \citep{Bartelmann:2001}. Therefore, $R_{A}$ can be expressed as \citet{Holanda:2016-2}
\begin{equation}\label{eq:RA1}
R_{A}(z_{l},z_{s})=\frac{D_{A_{ls}}}{D_{A_{s}}}=1-\frac{(1+z_{l})D_{A_{l}}}{(1+z_{s})D_{A_{s}}}.
\end{equation}
From equation (\ref{eq:RA1}), the ratio of $D_{A_{ls}}$ and $D_{A_{s}}$ can be converted to the ratio of $D_{A_{l}}$ and $D_{A_{s}}$, which can be further converted to the ratio of $D_{L_{l}}$ and $D_{L_{s}}$ using the DDR.

In the anisotropic universe described by Finsler geometry, the redshift is expressed as a dipolar structure \citep{Chang:2013b,Chang:2014,Li:2015}. Furthermore, the low multipole models have been developed to analyze the anisotropic cosmology \citep{Lineweaver:1996,Tegmark:2003,Copi:2010,Frommert:2010}. Therefore, considering the anisotropic cosmological models, we tentatively parameterize the DDR as a dipolar form
\begin{equation}\label{eq:DDR_cos}
\frac{D_{A}(1+z)^{2}}{D_{L}}=1+A{\rm cos}\theta,
\end{equation}
where $\theta$ is the angle between two objects for which $D_L$ and $D_A$ are measured (here it is the angle between SNe Ia and SL), $A$ is the anisotropic amplitude.

Combining equations (\ref{eq:RA1}) and (\ref{eq:DDR_cos}), we can obtain
\begin{equation}\label{eq:RA2}
R_{A}(z_{l},z_{s})=1-R_{L}(z_{l},z_{s})q,
\end{equation}
where
\begin{equation}\label{eq:q}
q\equiv\frac{\left(1+z_{s}\right)\left(1+A{\rm cos}\theta_{l}\right)}{\left(1+z_{l}\right)\left(1+A{\rm cos}\theta_{s}\right)},
\end{equation}
$R_{L}\equiv D_{L_{l}}/D_{L_{s}}$, $D_{L_{l}}$ ($D_{L_{s}}$) is the luminosity distance at the redshift $z_{l}$ ($z_{s}$), $\theta_{l}$ ($\theta_{s}$) is the angle between the SNe Ia and lens (source). The luminosity distance ratio $R_{L}$ can be extracted from the SNe Ia data according to equation (\ref{eq:mu}), and the angular diameter distance ratio $R_A$ can be extracted from the SL data according to equation (\ref{eq:RA}). By directly comparing $R_L$ and $R_A$, we can test whether equation (\ref{eq:RA2}) is valid or not. Note that although the lens and source are approximately at the same direction in the sky, the SNe Ia matched with lens and source are usually at different direction. Hence $\theta_l$ and $\theta_s$ are in general not the same. Due to this property, we can use the DDR to test the anisotropy of universe. Otherwise, if $\theta_l\equiv \theta_s$, the anisotropic signal is completely cancelled out in equation (\ref{eq:q}). The uncertainty of $R_{L}$ is obtained using the standard error propagation technique,
\begin{equation}
\triangle R_{L}=R_{L}\frac{{\rm ln}10}{5}\sqrt{\triangle \mu_{l}^{2}+\triangle \mu_{s}^{2}},
\end{equation}
where $\triangle \mu_{l}$ ($\triangle \mu_{s}$) is the uncertainty of distance modulus of SNe Ia matched with SL at the redshift of lensing (source), and $\ln$ is the natural logarithm.

\section{Samples and Results}\label{sec:Samples and Results}

Our SL sample consists of 118 strong lensing systems in the redshift range $z_l\in[0.075,1.004]$ for the lens and $z_s\in[0.196,3.595]$ for the source, compiled in \citet{Cao:2015}. As for the SNe Ia sample, we use the Union2.1 sample \citep{Suzuki:2012} consisting of 580 well-calibrated SNe Ia in the redshift range $z\in[0.015,1.414$], and JLA compilation \citep{Betoule:2014} involving 740 SNe Ia in the redshift range $z\in[0.01,1.30$].

It is clearly known from equation (\ref{eq:RA2}) that, for each SL system, two SNe Ia located at redshift $z_{l}$ and $z_{s}$ respectively should be found from SNe Ia. However, there is no SNe Ia that exactly locates at the redshit $z_{l}$ or $z_{s}$. To solve this problem, we apply the following approximation \citep{Holanda:2010,Yang:2013,Liao:2016}. We adopt a way to compare the redshift difference $\Delta z$ between the lens (source) and SNe Ia: i. If $\Delta z\leq 0.005$, the redshift difference is smaller enough to be ignored. ii. If there are two or more SNe Ia satisfying $\Delta z\leq 0.005$, we choose the one who has the smallest $\Delta z$. iii. To avoid correction, SNe Ia can't be used again when it is matched to other SL system. Using the Union2.1 dataset, this procedure results in 59 SL systems who have matched SNe Ia. The spatial distribution of the filtered SNe Ia and SL systems are plotted in the sky of equatorial coordinates in Fig.1. It should be noticed that the distance moduli of Union2.1 presented in \citet{Suzuki:2012} is calibrated in the $w$CDM model. To avoid the model dependence, in principle the original light-curve parameters should be used to recalibrate the data. However, the sample of the selected 59 pairs of SNe Ia is not large enough to make a tight constraint on the cosmological parameters, even for the simplest $\Lambda$CDM model. As \citet{Holanda:2016} pointed out, this dependence is much smaller than the errors of the gravitational lensing, hence can be ignored. On the other hand, the anisotropy of universe is very small and it can be regarded as a perturbation of the isotropic universe. Therefore, we directly use the distance moduli calibrated in the $w$CDM model and fit the filtered data to equation (\ref{eq:RA2}). The parameters are obtained using the least-$\chi^2$ method,
\begin{equation}\label{eq:chi-sl}
\chi^{2}_{\rm SL}=\sum^{N}_{i=1}\left[\frac{R_{A}(z_{l_{i}},z_{s_{i}})-1+q R_{L}(z_{l_{i}},z_{s_{i}})}{\sigma_{i}}\right]^{2},
\end{equation}
where $\sigma=\sqrt{\triangle R_{A}^{2}+q^{2}\triangle R_{L}^{2}+\sigma_{\rm int}^2}$ is the uncertainty propagated from the uncertainties of $R_L$, $R_A$ and some unidentified systematic uncertainties in the data. The free parameters are $A$, $f$ and $\sigma_{\rm int}$.

\begin{figure}\label{fig:Uniondirection}
\center
\includegraphics[width=0.6\textwidth]{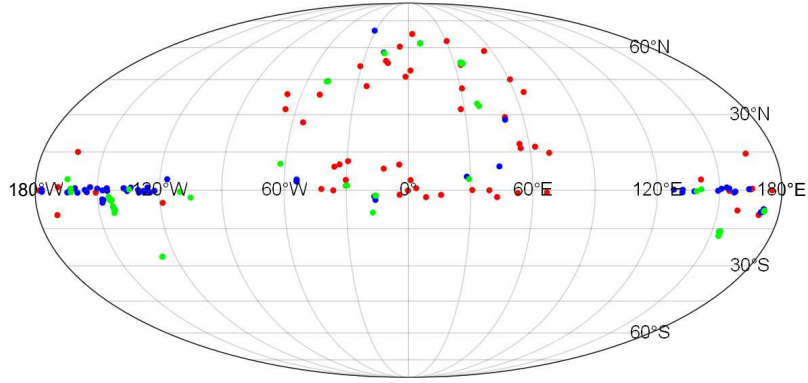}
\caption{\small{The red, blue and green dots denote the locations of SL systems matched with the Union2.1 compilation, SNe Ia located at $z_l$ and SNe Ia located at $z_s$, respectively.}}
\end{figure}

We list the best-fitting parameters and the reduced chi-square $\chi^{2}/{\rm dof}$ in the second column of Table \ref{tab:2.1}, where ${\rm dof}=N-p$ is the degree of freedom, $N$ is the number of data points and $p$ is the number of free parameters. We can see that the DDR is consistent with isotropy in the dipolar parametrization.

Additionally, the angular diameter distance can also be obtained from the galaxy cluster using the Sunyaev-Zeldovich effect. The derived angular diameter distance depends on the mass model of cluster. Two models are often used. One is the spherical symmetry model and the other is the ellipsoidal model. For the former there are 38 clusters presented in \citet{Bonamente:2006}, and for the later there are 25 clusters presented in \citet{De Filippis:2005}. Following \citet{Cao:2016}, we also add the galaxy clusters to SL systems to test DDR with dipolar structure. Using equation (\ref{eq:DDR_cos}), the distance modulus of cluster is given by
\begin{equation}\label{eq:mu_c}
\mu_{\rm cluster}(z)=5{\rm log}_{10}\left[\frac{D_{A_{\rm cluster}}(1+z)^{2}}{1+A{\rm cos}\theta_{c}} \right]+25,
\end{equation}
where $\theta_{c}$ is the angle between the cluster and SNe Ia matched at the same redshift $z$. We filter the SNe Ia data as above criteria, then fit the distance modulus of the filtered SNe Ia and cluster with the least-$\chi^{2}$,
\begin{equation}\label{eq:chi-c}
\chi^{2}_{\rm cluster}=\sum^{N}_{i=1}\left[\frac{\mu_{\rm cluster}(z_{i})-\mu_{\rm sn}(z_{i})}{\sigma_{i}}\right]^{2},
\end{equation}
where $\mu_{\rm sn}$ is the distance modulus of SNe Ia matched with galaxy cluster at the same redshift, \textbf{$\sigma=\sqrt{\triangle \mu_{\rm cluster}^{2}+\triangle \mu_{\rm sn}^{2}+\sigma_{\rm int}^2}$}.
In the case of the sample containing SL systems and galaxy clusters, the total $\chi^{2}$ is expressed by
\begin{equation}\label{eq:chi-c}
\chi^{2}=\chi^{2}_{\rm SL}+\chi^{2}_{\rm cluster}.
\end{equation}
The best-fitting parameters are obtained by minimizing equation (\ref{eq:chi-c}).

The results of adding galaxy clusters to SL systems are presented in the third and fourth columns of Table \ref{tab:2.1}, respectively.In the elliptical cluster model, the DDR is valid that is consistent with the pure SL systems. However, in the spherical cluster model, the DDR is violated more than $2\sigma$ significant.

\begin{table}
\caption{The best-fit dipolar parameters with the Union 2.1 sample. The angular diameter distance is derived from three cases: strong lensing (SL), strong lensing + elliptical cluster model (E), and strong lensing + spherical cluster model (S).}
\label{tab:2.1}
\centering
{\begin{tabular}{llll}
\hline\hline\noalign{\smallskip}
  &SL  & SL+25 Clusters(E)     & SL+38 Clusters(S)\\
\noalign{\smallskip}\hline\noalign{\smallskip}
$A$ &$0.038\pm0.063$   &$0.025\pm0.050$  &$-0.110\pm0.049$  \\
$f$ &$1.064\pm0.019$  &$1.062\pm0.019$  &$1.092\pm0.031$ \\
$\sigma_{\rm int}$ 	& 0.107  &  0.104 & 0.210 \\
$\chi^{2}$/dof  &1.000       &0.998         &0.997          \\
\noalign{\smallskip}\hline
\end{tabular}}
\end{table}

To compare different SNe Ia datasets, we also match SL systems with the JLA compilation \citep{Betoule:2014}. We use the similar criteria as in the Union2.1 case to match the JLA with SL systems. This results to 55 SL systems and the corresponding 55 pairs of SNe Ia. The spatial distribution of the filtered SNe Ia and SL systems are plotted in the sky of equatorial coordinates in Fig.2. The 55 pairs of SNe Ia sample filtered is insufficient to give a suitable constraint on the nuisance parameters in equation (\ref{eq:mu}). Here, we fix the parameters in equation (\ref{eq:mu}) to the values obtained by calibrating the full JLA compilation in the flat $\Lambda$CDM model, i.e. $M_{B} = -19.05$, $\alpha = 0.141$, $\beta = 3.101$, as is given by \citet{Betoule:2014}. We also add two sets of galaxy clusters to SL systems to give a combined constraint. The results are listed in Table \ref{tab:JLA}. For the pure SL sample, the discrepancy of DDR is at the level of more than 1$\sigma$. If we add the elliptical galaxy clusters, the DDR is consistent with isotropy, while the DDR is violated more than 1$\sigma$ if the spherical galaxy clusters are added.

\begin{figure}\label{fig:JLAdirection}
\center
\includegraphics[width=0.6\textwidth]{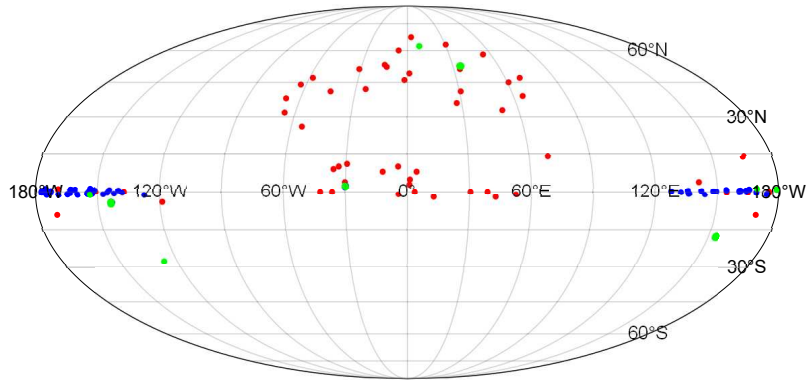}
\caption{\small{The red, blue and green dots denote the locations of SL systems matched with the JLA compilation, SNe Ia located at $z_l$ and SNe Ia located at $z_s$, respectively.}}
\end{figure}

\begin{table}
\caption{The best-fit dipolar parameters with the JLA sample. The angular diameter distance is derived from three cases: strong lensing (SL), strong lensing + elliptical cluster model (E), and strong lensing + spherical cluster model (S).}
\label{tab:JLA}      % Give a unique label
%\resizebox{!}{0.97cm}
\centering
\centering
{\begin{tabular}{llll}
\hline\hline\noalign{\smallskip}
  &SL  & SL+25 Clusters(E)     & SL+38 Clusters(S)\\
\noalign{\smallskip}\hline\noalign{\smallskip}
$A$ &$0.068\pm0.049$   &$0.026\pm0.044$  &$-0.064\pm0.046$  \\
$f$ &$1.057\pm0.018$  &$1.058\pm0.019$  &$1.076\pm0.027$ \\
$\sigma_{\rm int}$ 	&0.107   &0.112  &0.180  \\	
$\chi^{2}$/dof  &0.995        &0.996        &0.997          \\
\noalign{\smallskip}\hline
\end{tabular}}
\end{table}

\section{Monte Carlo Simulation}\label{sec:Simulation}
We apply the Monte Carlo (MC) simulation to test the statistical signification of our method. Here we take all the SL systems compiled in \citet{Cao:2015} as a sample. We firstly introduce the details about the mock dipolar data. For the supernova sample, we assume that there exist corresponding supernovas at $z_l$ and $z_s$ for all SL systems, and the distance modulus of supernova has a dipolar structure
\begin{equation}\label{eq:mu_dip}
\mu^{dip}(z)=\mu^{iso}(z)(1+D\cos\theta_p),
\end{equation}
where $\mu^{iso}(z)$ is the isotropic modulus calculated in a classical $\Lambda$CDM model with Hubble constant $H_0$  = 70 km s$^{-1}$ Mpc$^{-1}$ and the dimensionless matter density parameter $\Omega_M$ = 0.27, $D$ and $\theta_p$ are the dipolar amplitude and the angle between the object (supernova located at $z_l$ or $z_s$) and the fiducial preferred direction, respectively. In the simulation, we take the fiducial dipolar amplidute $D$ = 1.2$\times 10^{-3}$ and preferred direction ($l_0$, $b_0$) = (310.6$^{\circ}$, -13.0$^{\circ}$) in the galactic coordinates \citep{Lin:2016(b)}. Then the luminosity distance ratio is
\begin{equation}\label{eq:RL_dip}
R_L^{dip}(z_l,z_s)=10^{\frac{\mu^{dip}(z_l)-\mu^{dip}(z_s)}{5}},
\end{equation}
where $\mu^{dip}(z_l)$ and $\mu^{dip}(z_s)$ are the dipolar distance moduli obtained from equation (\ref{eq:mu_dip}). For SL systems, since the directions of lens and source are approximately the same, the anisotropic effects are canceled out in the angular diameter distances ratio $R_A$. Hence
\begin{equation}\label{eq:RA_dip}
R_A^{dip}=D^{dip}_{A_{ls}}/D^{dip}_{A_s}=D^{iso}_{A_{ls}}/D^{iso}_{A_s},
\end{equation}
where $D^{iso}_A$ is the isotropic angular diameter distance, defined with
\begin{equation}
D^{iso}_A(z)=\frac{c}{1+z}\int^z_0\frac{dz^{'}}{H(z^{'})},
\end{equation}
here $H(z)$ is the Hubble function. With regard to the uncertainties, we derive $\triangle R_{A}$ with equation ($\ref{eq:delta_RA}$) and $\triangle R_{L}$ by assuming $\triangle\mu=0.2$ for all supernovae. However, we find that the relative error has an impact on result. Comparing with the work of \citet{Lin:2016(b)}, the relative error in our work is $\triangle R_L/R_L^{dip}\gg \triangle\mu/\mu$. Hence, we shrink the uncertainties ten times i.e. $\triangle R_{L}^{dip}=\triangle R_L/10$ and $\triangle R_{A}^{dip}=\triangle R_A/10$.

We create a mock samples A to check whether our dipolar parametrization can test the anisotropic model in a real anisotropic dataset. In the sample A, the positions of SL systems are real, but the positions of mock supernova sample are homogeneously distributed in the sky. For $i$th SL system and corresponding supernovas, $R_L(z_{l_i},z_{s_i})$ is a random number generated from the Gaussian distribution $G(R_L^{dip}(z_{l_i},z_{s_i}),\triangle R_L^{dip}(z_{l_i},z_{s_i}))$, where $R_L^{dip}(z_{l_i},z_{s_i})$ is derived from equation (\ref{eq:RL_dip}). $R_A(z_{l_i},z_{s_i})$ is a random number generated from the Gaussian distribution $G(R_A^{dip}(z_{l_i},z_{s_i}),\triangle R_A^{dip}(z_{l_i},z_{s_i}))$, where $R_A^{dip}(z_{l_i},z_{s_i})$ is derived from equation (\ref{eq:RA_dip}). We replace the SL and SNe Ia matched in equation ($\ref{eq:RA2}$) with the sample A and use the least-$\chi^2$ method to search for the dipolar magnitude $A$. It is worth mentioning that the magnitude $A$ in this situation is a compositive effect of preferred direction and dipolar amplitude of distance modulus, which means that $A$ is not strictly equal to $D$. The results of dipolar amplitudes in 1000 MC simulations are plotted in Fig.3. The histogram can be well fitted by the Gaussian function with an average value $\bar{A}=(1.02\pm 0.01)\times 10^{-2}$ and standard deviation $\sigma_{A}=(0.44\pm0.01)\times 10^{-2}$. It indicates that, in a dipolar sample, the anisotropy can be verified at more than 2$\sigma$ confidence level.

\begin{figure}\label{fig:real_dip}
\center
\includegraphics[width=0.45\textwidth]{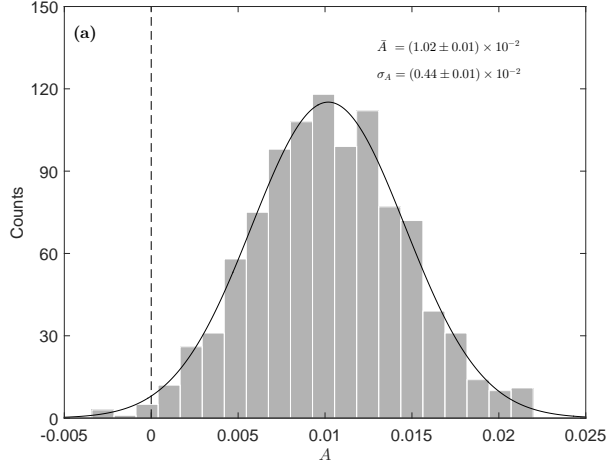}
\caption{\small{The histogram of dipolar amplitudes of mock sample A in 1000 MC simulations. The black curve is plotted by the best-fitting Gaussian function. The dashed vertical line represents the isotropy.}}
\end{figure}

Additionally, we construct another mock sample B to check whether our method would detect pseudo-anisotropy in an isotropic dataset. Compared with the sample A, the only difference in the sample B is that the distance modulus is isotropic, in other words, the amplitude $D$ = 0 in equation (\ref{eq:mu_dip}). The statistic results of dipolar amplitude $A$ in 1000 MC simulations are shown in Fig.4. It can be best fitted by Gaussian function with the best-fitting parameters $\bar{A}=(0.03\pm 0.11)\times 10^{-3}$ and $\sigma_{A}=(3.05\pm0.11)\times 10^{-3}$. This implies that there does not exist any violation of DDR in a fully isotropic sample.

\begin{figure}\label{fig:real_iso}
\center
\includegraphics[width=0.45\textwidth]{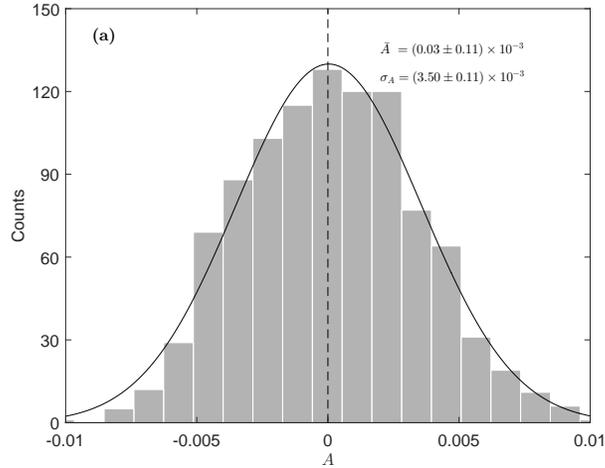}
\caption{\small{The histogram of dipolar amplitudes of mock sample B in 1000 MC simulations. The black curve is plotted by the best-fitting Gaussian function. The dashed vertical line represents the isotropy.}}
\end{figure}

If the anisotropic models were verified with our method, we would further find the preferred direction. Due to the disappearance of anisotropy in the angular diameter distance ratio, it is feasible to rotate all the SL data to a specific direction, which can be treated as the preferred direction. We create a mock sample C to search for the anisotropic amplitude and preferred direction. In the sample C, the SL sample also is real data compiled in \citet{Cao:2015}, but their directions all point to preferred direction parameterized as $\hat{n}$ = $(l, b)$ in the galactic coordinate, where $l\in [0^{\circ}, 360^{\circ}]$ and $b\in [-90^{\circ}, 90^{\circ}]$. The process of generating the ratios $R_A$ and $R_L$ is same as the sample A. In this situation, $\theta_l$ ($\theta_s$) in equation (\ref{eq:q}) now becomes the angle between the supernova located at $z_l$ ($z_s$) and preferred direction $\hat{n}$. The free parameters are anisotropic amplitude $A$ and two angular dimensions ($l$, $b$) characterizing $\hat{n}$. We apply the sample C to equation (\ref{eq:RA2}) and obtain the results of 1000 MC simulation with least-$\chi$ method, shown in Fig.5. The histogram of dipolar amplitude $A$ in panel (a) is well depicted by the Gaussian distribution with an average value $\bar{A}=(2.38\pm 0.01)\times 10^{-2}$ and standard deviation $\sigma_{A}=(3.60\pm0.08)\times 10^{-3}$. The anisotropy is embodied in the dipole of luminosity distance in our parametrization. The average dipolar amplitude is well consistent with the dipolar magnitude of luminosity distance in \citet{Chang:2014}. In panel (b), the dipolar direction of distance modulus and its antipode are respectively denoted with the black diamond pointing to ($l_0$, $b_0$) = (310.6$^{\circ}$, -13.0$^{\circ}$) and the black triangle. The gray dots, representing the mock dipolar directions in the sky of galactic coordinates, cluster near the black triangle i.e. antipole of the dipolar direction of distance modulus. It is about 94.4 percent of 1000 MC simulations that the mock directions distribute in a circular region of radius $\triangle\theta<20^{\circ}$ (denoted with the solid circle) centring on the antipole of the dipolar direction of distance modulus. It demonstrates that our method can accurately detect the dipolar amplitude and preferred direction in a fiducial dipolar dataset.

\begin{figure}
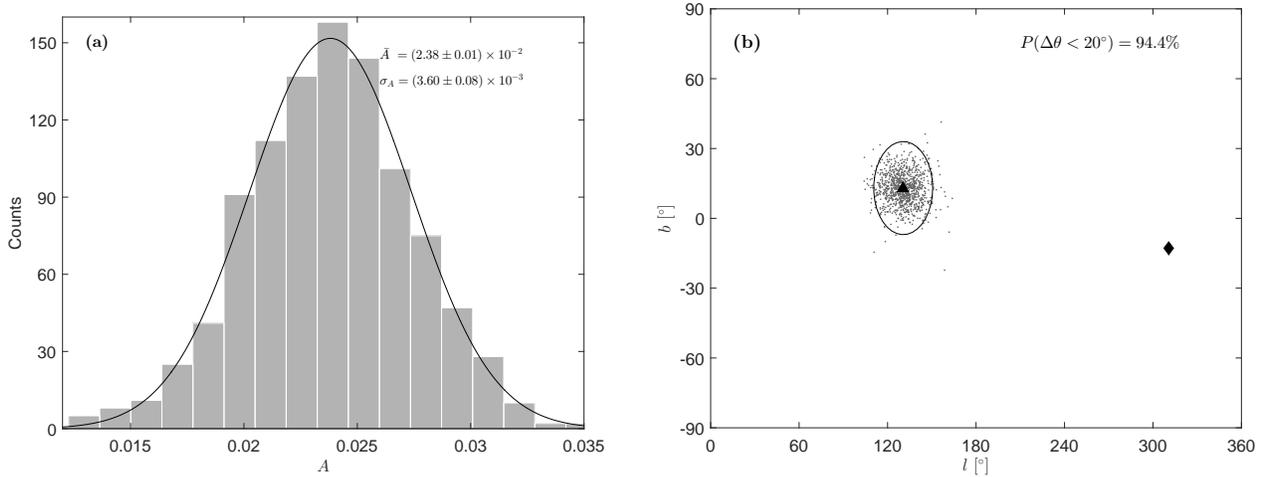
\label{fig:sim_dip}
\centering
\includegraphics[width=0.45\textwidth]{113SL_dip_sim.eps}\ \ \ \ \ \ \
\includegraphics[width=0.45\textwidth]{angle_dip_sim.eps}
\caption{\small{The results of mock sample C in 1000 MC simulations. Panel (a): the histogram of dipolar amplitudes with the black curve plotted by the best-fitting Gaussian function. Panel (b): the distribution of dipolar direction in the sky of galactic coordinates. The black diamond and black triangle respectively are the fiducial dipolar direction pointing to ($l_0$, $b_0$) = (310.6$^{\circ}$, -13.0$^{\circ}$) and its antipode. The solid circle denotes the circular region of radius $\triangle\theta<20^{\circ}$, centring on the antipode of the
fiducial direction.}}
\end{figure}

In addition, we construct a fiducially isotropic sample called sample D to cross check the method of finding the dipolar direction. Comparing with the sample C, the only difference is that the mock distance moduli of supernovae are isotropic, viz., that the dipolar amplitude $D$ = 0 in sample D. The results of 1000 MC simulations are depicted in Fig.6. In panel (a), the histogram of dipolar amplitudes can be best fitted to Gaussian function with the best-fitting parameters $\bar{A}=(4.90\pm 0.16)\times 10^{-3}$ and $\sigma_{A}=(2.60\pm0.17)\times 10^{-3}$. Because the probability that dipolar amplitudes are larger than that of sample C is near zero, the anisotropy detected in sample D can be treated as a pseudo-anisotropy caused by the statistical noise. Moreover, the distribution of the mock dipolar directions in panel (b) is expectedly homogeneous in the sky.

\begin{figure}
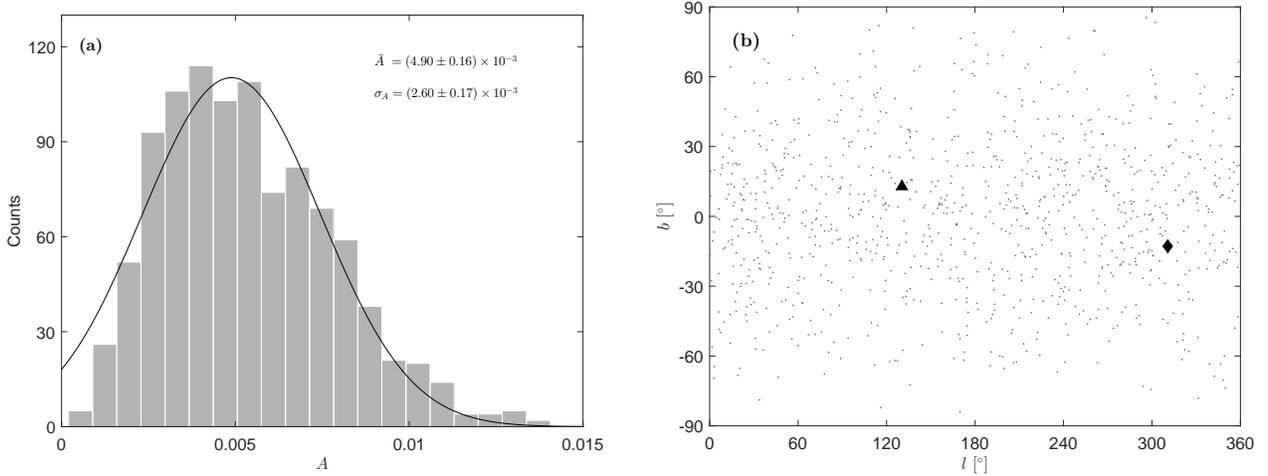
\label{fig:sim_iso}
\center
\includegraphics[width=0.45\textwidth]{113SL_iso_sim.eps}\ \ \ \ \ \ \
\includegraphics[width=0.45\textwidth]{angle_iso_sim.eps}
\caption{\small{The results of mock sample D in 1000 MC simulations. Panel (a): the histogram of dipolar amplitudes with the black curve plotted by the best-fitting Gaussian function. Panel (b): the distribution of dipolar direction in the sky of galactic coordinates. The black diamond and black triangle respectively are the fiducial dipolar direction pointing to ($l_0$, $b_0$) = (310.6$^{\circ}$, -13.0$^{\circ}$) and its antipode.}}
\end{figure}

In conclusion, our method of applying the DDR relation can effectively test the anisotropic cosmological models which are not based on the Riemannian spacetime. Additionally, once the anisotropy is detected with our method, we also can further search for the dipolar direction with the same method.

\section{Discussion and Summary}\label{sec:summary}

Any variation of the DDR would implies that there is new physics beyond the standard cosmological model. If the universe is depicted by the anisotropic cosmological models which are not based on the Riemannian spacetime, both the luminosity distance $D_{L}$ and the angular diameter distance $D_{A}$ are direction-dependent. In this paper, we combined the SNe Ia, SL systems and galaxy clusters to test the anisotropy with DDR. Differing from previous works, the directions of SNe Ia and SL systems (or galaxy clusters) should be considered in the anisotropic cosmology. We phenomenologically parameterized the DDR to a dipolar form, and constrained the anisotropic parameters using the combined dataset. The luminosity distance is measured from two different compilation of SNe Ia (Union2.1 and JLA), and the angular diameter distance is measured from SL systems and galaxy clusters.

The results presented in Table \ref{tab:2.1} within the Union2.1 compilation show that, for pure SL systems and elliptical cluster model added, the DDR validity is verified at 1$\sigma$. For the spherical cluster model, the DDR is violated at $\sim 2\sigma$ confidence level. Filtering SNe Ia data from the JLA compilation, the results are presented in Table \ref{tab:JLA}. In pure SL systems and spherical model added, the discrepancy of DDR is at more than 1$\sigma$ confidence level. While  the DDR is valid for the elliptical profile model. Moreover, we constructed the mock dipolar sample A and the mock isotropic sample B to test the signification of our method with the MC simulation. The results shown in Fig.3 and Fig.4 indicate that our method actually can test the anisotropic cosmological models in an anisotropic dataset. We also produced the mock dipolar sample C and the mock isotropic sample D to illustrate how to search for the dipolar direction with this method, if the universe were described by an anisotropic model. The results of MC simulations shown in Fig.5 and Fig.6 demonstrate that the dipolar direction of anisotropic sample with a fiducial dipolar direction can be correctly reproduced with our method.

The dipole model with a constant dipole amplitude used in our paper is the simplest parametrization of the anisotropy. Present cosmological observations do not find any sign of the anisotropy. Thus, any parameterizations should return to 1 at local Universe ($z = 0$). Therefore, we have investigated a redshift-dependant parametrization, i.e., $D_{A}(z)(1+z)^{2}/D_{L}(z)=1+A_0 z \cos\theta$. Using the new parametrization, the dipole amplitude $A_0$ constrained from SL+Union2.1 is $A_0=0.079\pm0.096$, and it is $A_0=0.190\pm0.087$ from SL+JLA. Similar to the old parametrization, from SL+Union2.1 the DDR still holds, but from SL+JLA the DDR is violated at about $2\sigma$. We can see that the dipole amplitude in the new parametrization is relatively larger than that in the old parametrization. This makes sense because the filtered SNe and SL data have an average redshift smaller than 1.

It should be noticed that our method is based on the flat Universe so that we can simplify the distance-sum-rule in Section \ref{sec:Observation and Methodology}. If the universe is not flat, the distance-sum-rule formula will complexly depends on the curvature and equations (\ref{eq:r}) and (\ref{eq:RA1}) will be more complex. Additionally, the current observations on CMB show that the spatial curvature $\Omega_k$ is very small \citep{Ade:2014,Ade:2016,Planck Collaboration:2018}, which is approximatively consistent with a flat Universe. Hence, for simplicity we just consider the flat case. Given the smallness of $\Omega_k$ and the large uncertainty of data, our results do not strongly depend on the curvature of the universe. With respect to data, the number of available SL system is no more than one half of the total SL systems due to the lack of matched SNe Ia. The available SL sample can be enlarged by adopting other techniques, such as using the polynomial fitting method to calculate the luminosity distance at any redshift, adding GRB data to the SNe Ia sample \citep{Holanda:2016-2}.

In summary, our method is statistically significant to test the anisotropic cosmological models which is not based on the  Riemannian spacetime. But due to the small data sample matched and large uncertainty, it is still premature to make a convincing conclusion in this paper.

\section*{Acknowledgments}
This work has been supported by the National Natural Science Fund of China under Grant Nos. 11775038, 11603005 and 11647307.

\bsp

\label{lastpage}

\end{document}